\documentclass[11pt]{article}
\usepackage{epsfig}

\title{\bf On the application of the
effective action approach to amplitudes with reggeon splitting}
\author{M.A.Braun, S.S.Pozdnyakov, M.Yu.Salykin, M.I.Vyazovsky\\
 Saint-Petersburg State University, Russia}

\def\beq{\begin{equation}}
\def\eeq{\end{equation}}
\def\tr{{\rm Tr}\,}
\def\pd{\partial}

\def\lra{\leftrightarrow}

\begin{document}
\maketitle
\medskip

{\bf Abstract}
Application of the effective action approach to amplitudes with loop
integration is studied for collisions on two and three centers with
possible gluon emission. A rule is formulated for the integration
around pole singularities in the induced vertices which brings the
results in agreement with the QCD. It is demonstrated that the amplitudes
can be restored from the purely transverse picture by introducing the
standard Feynman propagators for intermediate gluons and quarks.

\section{Introduction}
In the Regge kinematics, relevant for high-energy hadronic processes,
in the framework of the perturbative QCD, the amplitudes can be conveniently
constructed using the effective action proposed by L.N.Lipatov ~\cite{lipatov}.
In this formalism gluons at fixed
rapidities are described by the standard field
$V_{\mu}=-it^{a}V^{a}_{\mu}$.
Regions with essentially different rapidities are connected by the
reggeon field  $A^{y}_{{\mu}}=-it^{a}A^{ya}_{\mu}$ with
only non-zero longitudinal components $A_{+}$ and $A_{-}$, describing the
reggeized gluons.

The effective Lagrangian is local in rapidity and describes the
self-interaction of gluons at a given rapidity by means of the usual
QCD Lagrangian
${\cal L}_{QCD}$ and their interaction with reggeons.
It has the form \cite{lipatov}:
\beq
{\cal L}_{eff}={\cal L}_{QCD}(V^{y}_\mu+A^{y}_\mu)
+ 2 \tr\Big(({\cal A}_+(V^{y}_+ +A^{y}_+)-A^{y}_+)\pd^2_{\perp} A^{y}_-
+ ({\cal A}_-(V^{y}_- +A^{y}_-)-A^{y}_-)\pd^2_{\perp} A^{y}_+\Big),
\label{e1}
\eeq
where
\beq
{\cal A}_{\pm}(V_{\pm})=
\sum_{n=0}^{\infty}(-g)^nV_{\pm}(\pd_\pm^{-1}V_\pm)^n
=V_{\pm}-gV_{\pm}\pd_\pm^{-1}V_\pm+g^2V_{\pm}\pd_\pm^{-1}V_\pm
\pd_\pm^{-1}V_\pm+ -\dots \ .
\label{e2}
\eeq
The shift $V_\mu\to V_\mu+A_\mu$ with $A_\perp=0$ is done
to exclude direct gluon-reggeon transitions.
The reggeon fields are assumed to be subject to kinematical conditions
$\partial_{-}A_{+}=\partial_{+}A_{-}=0$
and their propagator is in the momentum representation
\beq
<A_+^{ya}A_-^{y'b}>=-i\frac{\delta_{ab}}{q_\perp^2}
\,\theta(y'-y) \ .
\label{e3}
\eeq

Inspection of Eq. (\ref{e2}) shows that the new vertices generated by the
effective action ("induced' vertices) contain poles at
$\partial _{\pm}=0$, which in the
momentum representation correspond to vanishing of the longitudinal momenta
transferred to the target or projectile. In fact these vertices can only
be introduced when these momenta are different from zero. Otherwise the
conditions of the reggeon kinematics are violated and the effective action
cannot be applied. Thus effective action serves only to find induced vertices
at non-zero values of the transferred longitudinal momenta. However, in
the physical amplitudes  these vertices are only a part of the whole
contribution. They are to be connected with the projectile(s) and target(s)
with reggeon propagators and in many cases integrated over the transferred
longitudinal momenta. At this moment the problem of interpreting the
mentioned singularities at zero values of these momenta arises.

Hermiticity of the effective Lagrangian suggests that from the start
the singularities at $\partial_{\pm}=0$ should possibly be interpreted in
the Cauchy principal value prescription in the momentum representation.
In our papers ~\cite{bravyaz,bralip} it was shown that for the
scattering on two centers with
gluon emission this prescription indeed produces correct amplitudes,
which in the lowest order reproduce the standard QCD amplitudes.
However, in a later paper M.Hentschinski discovered that for simple
elastic scattering on three centers the principal value prescription
for the effective Lagrangian violates the desired properties of the
transition vertex  of a reggeon into three reggeons ('R$\to$RRR vertex')
and in all probability to more reggeons ~\cite{hent1,hent2}. To restore
these properties M.Hentschinski proposed a recipe, which
essentially consists in projecting the contribution of the vertex
onto maximally antisymmetric colour
states in the crossed channel. Obviously this recipe is external to the
effective action approach and should be invoked as an additional
requirement. However, this recipe refers only to the vertex itself and
does not cover the case when the vertices are inserted into the amplitudes
as a whole and when it does not solve the problem.

In this paper we study this question in a more general framework and propose
a different prescription. In the effective action
the propagators of the projectile coupled to the reggeons should not be taken
as the standard Feynman ones. Only the $\delta$-functional part of them
should be retained in accordance with the fact that the effective action is
local in rapidity ~\cite{bralip}.
Note that the product of dropped parts with principal value
singularities may itself contain $\delta$-functional terms.
So the recipe in  ~\cite{bralip} cannot be formulated as
discarding all $\delta$-functional terms in the product of
intermediate projectile propagators as a whole. Rather such
terms should be dropped in each intermediate propagator.

We advocate that in accordance with
the Regge kinematics one should operate with the induced vertex as
if the transferred longitudinal momenta were different from zero.
We show that after its transformation into a certain adequate form
one can impose the prescription of principal value
for the singularities of the effective action in the longitudinal momenta.
The 'adequate form' is such that all the resulting propagators
have the standard Feynman singularities.
Note that taking the transferred longitudinal momenta different from zero
from the start we do not pretend to specify a description to circumvent
the singularities. The question of the correct pole prescription is
avoided at this point and in fact  postponed for later analysis of the
amplitude as a whole. In this analysis one indeed finds ambiguities as to
discarding the $\delta$-functional terms. They are resolved in a unique
manner by requiring that in the end one finds only the Feynman singularities
as dictated by comparison with the QCD diaframs.
We shall demonstrate that this rule allows to obtain
correct amplitudes from the effective action without any additional
non-physical terms. One can mnemonically term this prescription as a rule
that the induced vertices cannot contain any $\delta$-functions in
the transferred longitudinal momenta, although the exact meaning
of this rule is as presented above.

We also show that there exists an alternative way for the construction of
amplitudes in the Regge kinematics, which completely avoids the mentioned
singularity problem. Since many years ago,
starting from ~\cite{BFKL} and followed by ~\cite{bartels} it was shown that
their multiple discontinuities can be constructed in a purely transverse
picture containing certain effective vertices (Lipatov and Bartels vertices)
for real gluon emission. We demonstrate that the amplitudes themselves can be
restored from this transverse picture by connecting the particles
(gluon and quarks) by standard Feyman propagators. In some cases this method
proves to be simpler that the direct application of the effective action
(see e.g.~\cite{brasal}).

The paper is organized as follows. In Sections 2 and 3 we consider
scattering on two centers without (Section 2) and with (Section 3)
gluon emission. In this case the principal value prescription works.
A part of the material in these sections has been already published
in ~\cite{hent2} and ~\cite{bralip} and is included to have a general
view on the problem. The new results refer to the
description in terms of Lipatov and Bartels vertices connected by Feynman
propagators. In Sections 4 and 5 we study scattering on three centers
without (Section 4) and with (Section 5) gluon emission.
Again a part of the material reproduces results of
~\cite{BSPV} necessary for presentation  but most of it is new.

In our study we simplify the projectile and targets to be quarks
of momentum $k$ for the projectile and $l$ for each of the two or three
targets. We assume $k_-=l_+=k_\perp=l_\perp=0$ and work in the c.m.
system $k_+=l_-$. The colour indices of reggeons attached to the targets
and that of the emitted gluon are denoted as $b_1,b_2,b_3$ and $a$
respectively. To economize on notations we denote
the products of projectile quark colour matrices $t^{b_1}t^{b_2}t^{b_3}$
simply as $(123)$ and their trace as $[123]$.
Also the longitudinal momenta $q_{1-}$, $q_{2-}$ and $q_{3-}$ will be denoted
as 1,2 and 3 when this does not lead to confusion.
Pole at zero values of longitudinal momenta will always be
understood in the principal value sense, which will not be indicated
explicitly. In our figures the normal gluons will be
shown by solid lines and reggeons by wavy lines. The quark projectile
will be shown by thick solid line. The induced vertices will be denoted by
open circles with cross, the effective vertices of Lipatov and Bartels by
simple and double dots.

\section{Elastic scattering off two centers}
\subsection{Lowest order QCD}
\begin{figure}[h]
\leavevmode \centering{\epsfysize=0.15\textheight\epsfbox{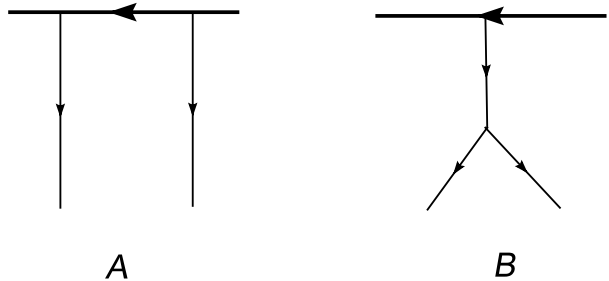}}
\caption{Elastic scattering on two centers in the lowest order of the QCD.
Here and in the following figures coupling of the reggeons to the target
quarks is not explicitly shown
}
\label{fig1}
\end{figure}

In the QCD in the lowest order the amplitude for the elastic
scattering off two centers in the axial gauge $(Vl)=0$ is trivial:
it is just the double gluon exchange, Fig. \ref{fig1},$A$,
since the contribution of the diagram with the 3-gluon vertex
Fig. \ref{fig1},$B$ is zero.
So the amplitude is
\beq
A=16(kl)^2\frac{(21)}{(k-q_1)^2+i0}+P_{12}\ .
\label{eq1}
\eeq
Here, as mentioned in the Introduction, $(21)=t^{b_2}t^{b_1}$;
$q_1$ is the momentum transferred to the right target.
In the high-energy limit $q_{1+}\to 0$. Symbol $P_{12}$ means
interchange $1\lra 2$. From the mass-conditions for the projectile
it follows that $q_{1-}+q_{2-}\to 0$. The Regge kinematics
assumes $|2k_{+}q_{1-}|>> |q_{1\perp}^2|$.
Then $(k-q_1)^2=-2k_{+}q_{1-}+q_{1\perp}^2\simeq -2k_{+}q_{1-}$.
In this limit the amplitude can be split into the principal value part
and $\delta$-functional part as follows
\beq
A=16(kl)^2\Big[-if^{b_1b_2c}t^c {\rm P}\frac{1}{2k_+q_{1-}}
-\pi\delta(2k_+q_{1-})\{t^{b_1}t^{b_2}\}\Big]+P_{12}\ .
\label{eq2}
\eeq

\subsection{Effective action result}
\begin{figure}[h]
\leavevmode \centering{\epsfysize=0.15\textheight\epsfbox{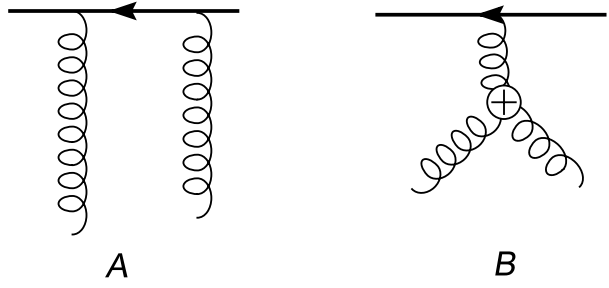}}
\caption{Elastic scattering on two centers in the effective action approach}
\label{fig2}
\end{figure}

In the effective action approach one has to retain only the $\delta$-
functional part of the diagram in Fig. \ref{fig2},$A$ but take into account
the diagram in Fig. \ref{fig2},$B$. Obviously to have the correct
amplitude the latter has to reproduce the term with the principal value in
(\ref{eq2}).
The standard 3-gluon vertex, as before, gives zero in the
axial gauge and only the induced vertex remains. This reggeon $\to$
two-reggeons vertex is given by
\beq
\Gamma_{R\to RR}=\frac{(q_1+q_2)^2_\perp}{q_{1-}}f^{b_1b_2c}\ .
\eeq
Inserting it into the amplitude in Fig. \ref{fig2},$B$ we obtain
the contribution
\beq
A_1=
-i16(kl)^2f^{b_1b_2c}t^c \frac{1}{2k_+q_{1-}}\ .
\label{eq3}
\eeq
Comparison with
Eq. (\ref{eq2}) demonstrates that if one understand the pole at
$q_{1-}=0$ in Eq. (\ref{eq3}) in the principal value sense
then the effective action exactly reproduces the QCD result.

Note that from this result it follows that one can describe the scattering
just by taking only diagrams A with the double reggeon exchange. In this way
one avoids poles at $q_{1-}=0$ altogether and remains with the standard
Feynman denominators. This presets a simple example of the alternative
description of high-energy amplitudes in the Regge kinematics.

\section{Gluon emission off two centers}
\subsection{Lowest order QCD}
\begin{figure}[h]
\leavevmode \centering{\epsfysize=0.25\textheight\epsfbox{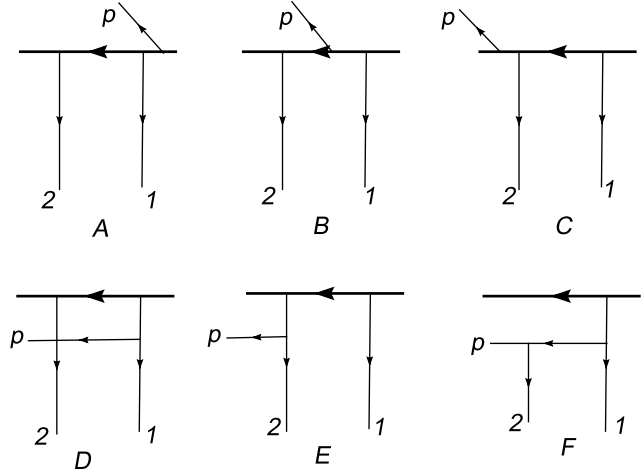}}
\caption{Gluon emission on two centers in the lowest order of the QCD}
\label{fig3}
\end{figure}

In the QCD at the lowest order it is described by 6 diagrams shown in
Fig. \ref{fig3},$A-F$.
These diagrams in the Regge kinematics
were calculated in ~\cite{bralip}. It was found
\beq
A+B+C=-32i(kl)^2\frac{(ep)_\perp}{p_\perp^2}
\Big[\frac{f^{b_1ac}(2c)}{(k'+q_2)^2+i0}-
\frac{f^{ab_2c}(c1)}{(k-q_1)^2+i0}\Big]\ ,
\label{dabc}
\eeq
\beq
D=-32i(kl)^2(e,p+q_1)_\perp\frac{f^{ab_1c}(2c)}{[(k'+q_2)^2+i0]
[(p+q_1)^2+i0]}\ ,
\label{dd}
\eeq
\beq
E=-32i(kl)^2(e,p+q_2)_\perp\frac{f^{ab_2c}(c1)}{[(k-q_1)^2+i0]
[(p+q_2)^2+i0]}\ ,
\label{de}
\eeq
\beq
F=32(kl)^2\frac{p_+}{k_+}\frac{(e,p+q_1+q_2)_\perp}{(p+q_1+q_2)_\perp^2}
\frac{f^{ab_2c}f^{cb_1d}t^d}
{(p+q_2)^2+i0}\ .
\label{df}
\eeq
Here $e$, $p$ and $a$ are the polarization vector, momentum and colour index
of the emitted gluon; $k'=k-q_1-q_2-p$; $(2c)=t^{b_2}t^c$.
To these diagrams also the ones with interchange $1\lra 2$ should be added.

\subsection{Effective action results}
\begin{figure}[h]
\leavevmode \centering{\epsfysize=0.25\textheight\epsfbox{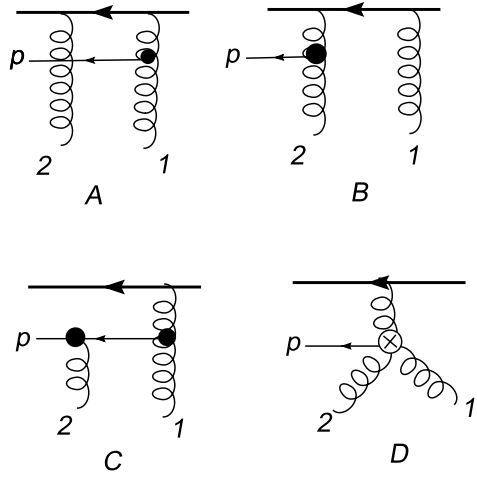}}
\caption{Gluon emission on two centers in the effective action approach}
\label{fig4}
\end{figure}

In the effective approach the same amplitude is described by only
four diagrams shown in Fig. \ref{fig4},$A-D$.
In  diagrams $A$ and $B$
the gluon is emitted by the Lipatov vertex.
\beq
L(p,q_1)=\frac{(pe)_\perp}{p_\perp^2}-
\frac{(p+q_1,e)_\perp}{(p+q_1)_\perp^2}\ .
\label{lv}
\eeq
According to the rules of the
effective action approach  in these diagrams in the quark propagator only
the $\delta$-functional term is to be retained.

The $R\to RRP$ vertex ($P$ for "particle') which enters diagrams C and D was
calculated in ~\cite{bravyaz}. It can be split into the proper (W) and
induced (R) parts
\beq
\Gamma_{R\to RRP}=W_{R\to RRP}+R_{R\to RRP} \ ,
\eeq
\beq
W_{R\to RRP}=-i\frac{2q_+q_\perp^2}{(q_2+p)^2+i0}f^{db_1c}f^{cb_2a}B(p,q_2,q_1),
\label{g1}
\eeq
\beq
R_{R\to RRP}=i\frac{q_\perp^2}{q_{1-}}f^{db_1c}f^{cb_2a}L(p,q_2).
\label{g2}
\eeq
Here $B(p,q_2,q_1)=L(p+q_2,q_1)$ is the so-called Bartels vertex.

Inserted into the amplitude the proper part W does not give
any trouble after integrations. However, in the induced part R we
again meet with the pole at $q_{1-}=0$. In ~\cite{bralip} it was demonstrated
that summed with the $\delta$-functional term from the quark propagator
the contribution from the $R\to RRP$ vertex contained in the part R
exactly reproduces the QCD result provided one interprets the pole in
(\ref{g2}) as the principal value. The necessity to remove $\delta$-functional
terms from the induced vertex does not arise.

However, this is not so for scattering on three (and possible more)
centers.

\subsection{Alternative form of the amplitude with gluon emission}
Here we present a different form of the amplitude in terms of the
Lipatov and Bartels vertices, which does not involve poles at
$q_{1,2-}=0$ and so does not require any additional information about
these poles.

In the Regge kinematics
\beq
\frac{1}{(p+q_1)^2+i0}=\frac{1}{-2p_+q_{2-}+(p+q_1)_\perp^2+i0}=
\frac{1}{(p+q_1)_\perp^2}\Big(1+\frac{2p_+q_{2-}}{(p+q_1)^2+i0}\Big)\ .
\eeq
Correspondingly we split contribution $D$, Eq. (\ref{dd}) to the amplitude
in two parts: $D=D_1+D_2$, where
\beq
D_1=-32i(kl)^2\frac{(e,p+q_1)_\perp}{(p+q_1)_\perp^2}
\frac{f^{ab_1c}t^{b_2}t^c}{(k'+q_2)^2+i0}\ ,
\label{dd1}
\eeq
\beq
D_2=-32i(kl)^2\frac{(e,p+q_1)_\perp}{(p+q_1)_\perp^2}
f^{ab_1c}t^{b_2}t^c\frac{2p_+q_{2-}}{[(k'+q_2)^2+i0][(p+q_1)^2+i0]}\ .
\label{dd2}
\eeq
Similarly we split contribution $E$, Eq. (\ref{de}) in two parts
\beq
E_1=-32i(kl)^2\frac{(e,p+q_2)_\perp}{(p+q_2)_\perp^2}
\frac{f^{ab_2c}t^{c}t^{b_1}}{(k-q_1)^2+i0}\ ,
\label{de1}
\eeq
\beq
E_2=-32i(kl)^2\frac{(e,p+q_2)_\perp}{(p+q_2)_\perp^2}
f^{ab_2c}t^{c}t^{b_1}\frac{2p_+q_{1-}}{[(k-q_1)^2+i0]
[(p+q_2)^2+i0]}\ .
\label{de2}
\eeq
Terms $D_1$ and $E_1$ summed with $A+B+C$ give
\beq
A+B+C+D_1+E_1=-32i(kl)^2 \left(
f^{b_1ac}t^{b_2}t^c\frac{L(p,q_1)}{(k'+q_2)^2+i0}
+ f^{b_2ac}t^ct^{b_1}\frac{L(p,q_2)}{(k-q_1)^2+i0} \right) ,
\eeq
which corresponds to diagrams Fig. \ref{fig4},$A,B$ with normal
Feynman propagators.

Interchange $(1\lra 2)$ in $D_2$ and $E_2$ gives contributions $\tilde{D}_2$
and $\tilde{E_2}$. One finds
\beq
D_2+\tilde{E}_2=32(kl)^2\frac{(e,p+q_1)_\perp}{(p+q_1)_\perp^2}
f^{ab_1c}f^{b_2cd}t^d\frac{p_+}{k_+}\frac{1}{(p+q_1)^2+i0}
\eeq
and $\tilde{D}_2+E_2$ is obtained after interchange $(1 \lra 2)$.
Summing this with $F$, Eq. (\ref{df}) and $\tilde{F}$ obtained after
$(1\lra 2)$ we get
\beq
F+\tilde{D}_2+E_2=32(kl)^2 f^{ab_2c}f^{b_1cd}t^d\frac{p_+}{k_+}
\frac{B(p,q_2,q_1)}{(p+q_2)^2+i0}\ ,
\eeq
\beq
\tilde{F}+D_2+\tilde{E}_2=32(kl)^2 f^{ab_1c}f^{b_2cd}t^d\frac{p_+}{k_+}
\frac{B(p,q_1,q_2)}{(p+q_1)^2+i0}\ .
\eeq
These contributions correspond to the diagram in Fig. \ref{fig4},$C$
and the one with $(1\lra 2)$ with the Bartels vertex and normal Feynman
propagator.

As a result the entire amplitude can be obtained from only diagrams
in Fig. \ref{fig4},$A, B$ and $C$ with the standard Feynman propagators
without the induced contribution in Fig. \ref{fig4},$D$. In this way,
as before, the problem of singularities at $q_{1,2-}=0$ does not arise
at all.

\section{Elastic scattering off three centers}
\subsection{Lowest order QCD}
\begin{figure}[h]
\leavevmode \centering{\epsfysize=0.15\textheight\epsfbox{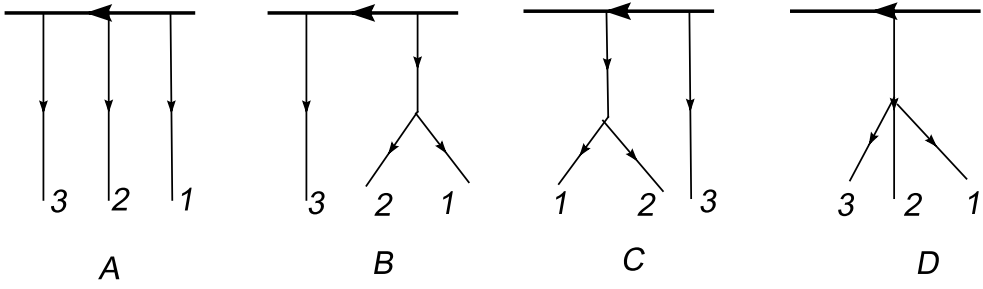}}
\caption{Elastic scattering on three centers in the lowest order of the QCD}
\label{fig5}
\end{figure}

In the axial gauge the amplitude for the elastic
scattering off three centers in the lowest order is again trivial
and reduces to the triple gluon exchange, Fig. \ref{fig5},$A$, since
diagrams \ref{fig5},$B-D$ with
both 3- and 4- gluon vertices give zero.
So the amplitude is
\beq
A=64i(kl)^3\frac{(321)}{[(k-q_1-q_2)^2+i0]
[(k-q_1)^2+i0]}+P_{123}\ ,
\label{eq4}
\eeq
where $P_{123}$ means adding contributions from the permutation of
(1,2,3). We recall that $(123)= t^{b_1}t^{b_2}t^{b_3}$.
Denoting $q_{1-}, q_{2-}$ and $q_{3-}$ simply as 1, 2
and 3 respectively, we rewrite (\ref{eq4}) as
\beq
A=i\frac{64(kl)^3}{4k_+^2}\frac{(321)}{[-(1+2)+i0]
[-1+i0]}+P_{123}\ .
\label{eq5}
\eeq

For the following it is useful to split the propagators into
the principal value and $\delta$-functional parts
\[
\frac{1}{[-(1+2)+i0][-1+i0]}=\Big(-\frac{1}{1+2}-i\pi\delta(1+2)\Big)
\Big(-\frac{1}{q_{1-}}-i\pi\delta(1)\Big)\]\beq=
\frac{1}{(1+2)1}+i\pi\frac{1}{q_{2-}}\Big(\delta(1)-\delta(3)\Big)-
\pi^2\delta(1)\delta(2).
\label{eq6}
\eeq
Here we used that in the Regge kinematics $1+2+3=0$.

In summing over permutations of $(1,2,3)$ it is convenient to combine
terms with order $(123)$ and $(321)$, since
\[2+3=-1,\ \ 3=-(1+2),\ \ 3\lra 1 , \]
so that the term with $(123)\to (321)$ will have the same real part
as (\ref{eq6}) but the imaginary part with the opposite sign.
Introducing
\[(123)_\pm=\frac{1}{2}\Big((123)\pm(321)),\]
we find
\[
A=i\frac{64(kl)^3}{2k_+^2}\Big\{
(321)_+\Big(\frac{1}{(1+2)1}-\pi^2\delta(1)\delta(2)\Big)
+(213)_+\Big(\frac{1}{(1+3)3}-\pi^2\delta(1)\delta(3)\Big)
\]\[+
(132)_+\Big(\frac{1}{(2+3)2}-\pi^2\delta(2)\delta(3)\Big)
+i\pi(321)_-\frac{1}{q_{2-}}\Big(\delta(1)-\delta(3)\Big)
\]\beq
+i\pi(213)_-\frac{1}{q_{1-}}\Big(\delta(3)-\delta(2)\Big)
+i\pi(132)_-\frac{1}{q_{3-}}\Big(\delta(2)-\delta(1)\Big)\Big\}.
\label{eq7}
\eeq

\subsection{Effective action result}
\begin{figure}[h]
\leavevmode \centering{\epsfysize=0.15\textheight\epsfbox{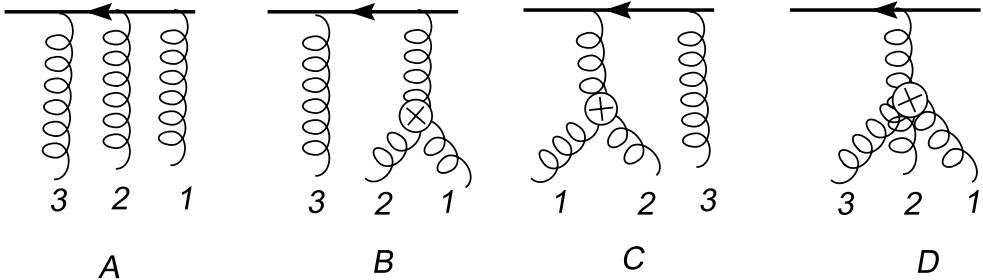}}
\caption{Elastic scattering on three centers in the effective action approach}
\label{fig6}
\end{figure}

In the effective action approach, apart from the triple reggeon exchange
we have additionally diagrams with induced 3-reggeon and 4-reggeon vertices
Fig. \ref{fig6},$B-D$.

In the triple gluon exchange we have to retain in the quark propagators
only the $\delta$-functional terms. Their contribution will obviously
reproduce terms with the product of two $\delta$-functions in (\ref{eq7}).

Let us start with diagram $D$ in Fig. \ref{fig6}. The induced 4-reggeon vertex
is
\beq
\Gamma_{R\to RRR}=2i g^2\frac{q^2_\perp}{(q_1+q_2)_-q_{1-}}
{\rm Tr}(t^at^{b_3}t^{b_2}t^{b_1})+P_{123}\ ,
\label{grrrr}
\eeq
where $q=q_1+q_2+q_3$. We denote
\[{\rm Tr}(t^at^{b_3}t^{b_2}t^{b_1})\equiv [a321]\]
and rewrite (\ref{grrrr}) as
\beq
\Gamma_{R\to RRR}=2i g^2\frac{q^2_\perp}{(1+2)1}[a321]+P_{123}\ .
\label{grrrr1}
\eeq
Combining terms with order $(123)$ and $(321)$ and assuming that
the denominators do not vanish we get
\beq
\Gamma_{R\to RRR}=4i g^2q^2_\perp\Big\{[a321]_+\frac{1}{(1+2)1}
+[a213]_+\frac{1}{(1+3)3}+[a132]_+\frac{1}{(3+2)2}\Big\},
\label{grrrr2}
\eeq
where $[a132]_+=(1/2)([a132]+[a231])$.

Inserting this into the amplitude we get additional
factor $16k_+l_-^3t^a /q^2_\perp$
with summation over $a$.
Presenting
$(123)=C^{123}+C^{d123}t^d$ , multiplying  and taking traces we find
\[[123]=N_cC^{123}=\frac{1}{4}(d^{123}+if^{123}),\ \
[a123]=\frac{1}{2}C^{a123}.\]
Thus
\[(123)=\frac{1}{4N_c}(d^{123}+if^{123})+2t^a[a123],\]
from which we find
\beq
t^a[a123]=\frac{1}{2}(123)-\frac{1}{8N_c}(d^{123}+if^{123}),\ \
t^a[a123]_+=\frac{1}{2}(123)_+-\frac{1}{8N_c}d^{123}.
\label{ta123}
\eeq

So we get for  diagram $D$ in Fig. \ref{fig6}
\beq
D=32 ik_+l_-^3\Big\{(321)_+\frac{1}{(1+2)1}
+(213)_+\frac{1}{(1+3)3}+(132)_+\frac{1}{(3+2)2}\Big\}+\Delta D ,
\label{eq8}
\eeq
where
\beq
\Delta D=32 ik_+l_-^3\frac{d^{123}}{8N_c}
\Big\{
\frac{1}{(1+2)1}+\frac{1}{(1+3)3}+\frac{1}{(3+2)2}\Big\}
=
-32 ik_+l_-^3\frac{d^{123}}{8N_c}
\Big\{\frac{1}{1\cdot 3}+\frac{1}{2\cdot 3}+\frac{1}{1\cdot 2}\Big\}.
\label{eq9}
\eeq
As mentioned in the Regge kinematics $1+2+3=0$. So if according to our
rules we take all $q_{i-}$, $i=1,2,3$, different from zero then
the r.h.s of Eq.(\ref{eq9}) vanishes.
In the principal value presription taken from the start
\beq
\frac{1}{1\cdot 3}+\frac{1}{2\cdot 3}+\frac{1}{1\cdot 2}=
-\pi^2\delta(1)\delta(2) .
\label{eq91}
\eeq
So in this case the induced vertex contains a $\delta$-functional
contribution.  Our rule is equivalent to dropping it. So with our rule
$\Delta D=0$ and comparing with (\ref{eq7}) we conclude that the
effective action correctly reproduces the part of the QCD amplitude
with the product of two principal value poles.

Note that without our rule $\Delta D$ is different from zero
and we get an additional contribution with the product
$\delta(1)\delta(2)$, which spoils the agreement with the QCD result.

Now to the diagrams $B$ and $C$. Inserting the vertex $\Gamma_{R\to RR}$
into the amplitude we get the product
\[it^{b_3}t^af^{ab_1b_2}=t^{b_3}[t^{b_1},t^{b_2}].\]
So we find
\beq
B=16(kl)l_-^2\pi\Big\{\delta(3)\frac{1}{q_{1-}}\Big((312)-(321)\Big) +
\delta(2)\frac{1}{q_{3-}}\Big((231)-(213)\Big)+
\delta(1)\frac{1}{q_{2-}}\Big((123)-(132)\Big)\Big\}.
\eeq
In the same way we find
\beq
C=16(kl)l_-^2\pi\Big\{\delta(3)\frac{1}{q_{1-}}\Big((123)-(213)\Big) +
\delta(2)\frac{1}{q_{3-}}\Big((312)-(132)\Big)+
\delta(1)\frac{1}{q_{2-}}\Big((231)-(321)\Big)\Big\}.
\eeq
In the sum \[B+C=\]
\beq
32 (kl)l_-^2\pi
\Big\{(312)_-\Big(\delta(3)\frac{1}{q_{1-}}+\delta(2)\frac{1}{q_{3-}}\Big)+
(123)_-\Big(\delta(3)\frac{1}{q_{1-}}+\delta(1)\frac{1}{q_{2-}}\Big)+
(231)_-\Big(\delta(2)\frac{1}{q_{3-}}+
\delta(1)\frac{1}{q_{2-}}\Big)\Big\}.
\eeq
Taking into account that $1+2+3=0$ we can rewrite it as
\beq
B+C=32 (kl)l_-^2\pi
\Big\{(312)_-\frac{1}{q_{1-}}\Big(\delta(3)-\delta(2)\Big)+
(123)_-\frac{1}{q_{2-}}\Big(\delta(1)-\delta(3)\Big)+
(231)_-\frac{1}{q_{3-}}\Big(\delta(2)-\delta(1)\Big)\Big\}.
\eeq
Comparing with (\ref{eq7}) we conclude that we reproduce the QCD result
with the pole singularities in the principal value
prescription.

So in the end we have demonstrated that the principal value prescription
for the effective action gives the correct amplitude for the
elastic scattering on three centers provided one drops $\delta$-functional
terms from the induced vertices and retains only those of them
that come from the projectile quark propagators.

Note that as for the elastic scattering off two centers we can forget about the
induced vertices and use only the triple reggeon exchange with full
Feynman propagators for the projectile quark, thus completely avoiding
discussion about pole singularities in $q_{i-}$, $i=1,2,3$.

\section{Gluon emission off three centers}
\subsection{QCD results}
{\bf 1. Three interactions of the projectile and the gluon emitted from it}

\begin{figure}
\hspace*{2.5 cm}
\begin{center}
\includegraphics[scale=0.75]{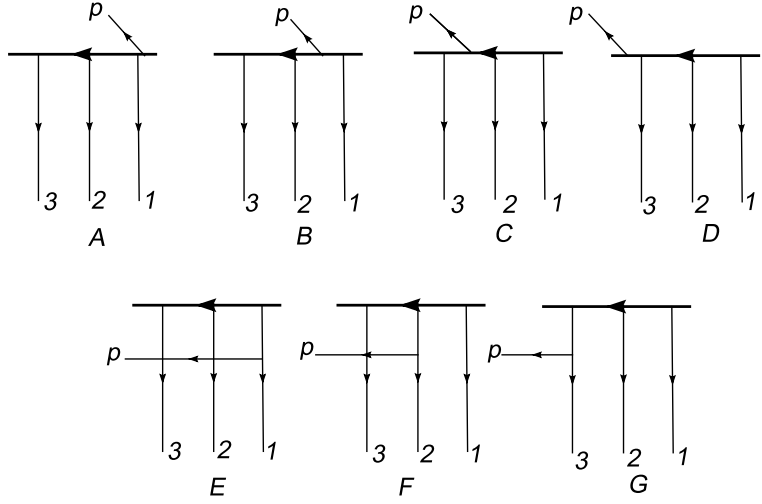}
\end{center}
\caption{Three interactions of the projectile with gluon emission}
\label{fig7}
\end{figure}

We have 4 diagrams shown in Fig \ref{fig7},$A-D$.
They all have a common factor $F_1$:
\[F_1=16i\frac{(kl)^3}{8k_+^3}\frac{k_+}{p_+}(ep)_\perp\]
($i^7$ from the quark line, $(-i)^3$ from interactions, $(-i)$ for
the amplitude and a minus from $(ek)$).

Now for the rest.
To have a more symmetric notations we denote $p_-\equiv q_{4-}$ and
$a\equiv b_4$. Then we find, neglecting the transverse parts
\[(7,A)=\frac{(3214)}{(-4+i0)(-(4+1)+i0)(-(4+1+2)+i0)}=
\frac{(3214)}{(-4+i0)(-(4+1)+i0)(3+i0)}\ ,\]
where we have used $1+2+3+4=0$.
Next
\[(7.B)=\frac{(3241)}{(-1+i0)(-(4+1)+i0)(3+i0)}\ ,\]
\[(7.C)=\frac{(3421)}{(-1+i0)(-(1+2)+i0)(3+i0)}\]
and finally
\[(7.D=\frac{(4321)}{(-1+i0)(-(1+2)+i0)(4+i0)}\ .\]

We transform
\[\frac{1}{(-1+i0)(-(1+4)+i0)}=\frac{1}{-4+i0}\Big(\frac{1}{-q_{1-}+i0}-
\frac{1}{-(q_1+p)_-+i0}\Big)\]
and correspondingly split (7.B) into two parts
$(7.B)=(7.B1)+(7.B2)$,
where
\[(7.B1)=\frac{(3241)}{(-4+i0)(-1+i0)(3+i0)}\ ,
\ \ (7.B2)=-\frac{(3241)}{(-4+i0)(-(1+4)+i0)(3+i0)}\ .\]
Similarly $(7.C)=(7.C1)+(7.C2)$ with
\[
(7.C1)=\frac{(3421)}{(4+i0)(-1+i0)(3+i0)}\ ,
\ \ (7.C2)=-\frac{(3421)}{(4+i0)(-1+i0)(3+4+i0)}\ .
\]

Now we combine (7.A) with (7.B2), (7.D) with (7.C2) and (7.B1) with (7.C1)
taking into account that $4=p_-$ does not vanish.
We get
\beq (7.A+B+C+D)
=\frac{1}{p_-}\Big\{\frac{i(32d)f^{41d}}{(-(1+4)+i0)(3+i0)}+
\frac{i(d21)f^{43d}}{(-1+i0)(3+4+i0)}+\frac{i(3d1)f^{42d}}
{(-1+i0)(3+i0)}\Big\}.
\label{emitq}
\eeq

{\bf 2. Three interactions of the projectile and the gluon emitted
from one of the interactions}

We have three  diagrams shown in Fig. \ref{fig7},$E-G$.
The common momentum factor is
\[F_2=-16\frac{(kl)^3}{4k_+^2}.\]
The rest gives
\[ (7.E+F+G)=\frac{[e(p+q_1)]_\perp}{(p+q_1)^2+i0}\,
\frac{(32d)f^{d41}}
{(-(1+4)+i0)(3+i0)}\]\[+
\frac{[e(p+q_2)]_\perp}{(p+q_2)^2+i0}\,
\frac{(3d1)f^{d42}}
{(-1+i0)(3+i0)}+
\frac{[e(p+q_3)]_\perp}{(p+q_3)^2+i0}\,
\frac{(d21)f^{d43}}
{(-1+i0)(3+4+i0)}\ .\]

Summing with the contributions from the previous section we find the
total contributions from three
interactions of the projectile as
\[ (7)=4\frac{(kl)^3}{k_+^2}\Big\{
\Big(\frac{(ep)_\perp}{p_\perp^2}-
\frac{[e(p+q_1)]_\perp}{(p+q_1)^2+i0}\Big)
\frac{(32d)f^{41d}}{(-(1+4)+i0)(3+i0)}
\]\[+
\Big(\frac{(ep)_\perp}{p_\perp^2}-
\frac{[e(p+q_2)]_\perp}{(p+q_2)^2+i0}\Big)
\frac{(3d1)f^{42d}}{(-1+i0)(3+i0)}
\]
\beq
+\Big(\frac{(ep)_\perp}{p_\perp^2}-
\frac{[e(p+q_3)]_\perp}{(p+q_3)^2+i0}\Big)
\frac{(d21)f^{43d}}{(-1+i0)(3+4+i0)}\Big\}\ .
\label{eq10}
\eeq

{\bf 3. Two interactions of the projectile}

\begin{figure}
\hspace*{2.5 cm}
\begin{center}
\includegraphics[scale=0.75]{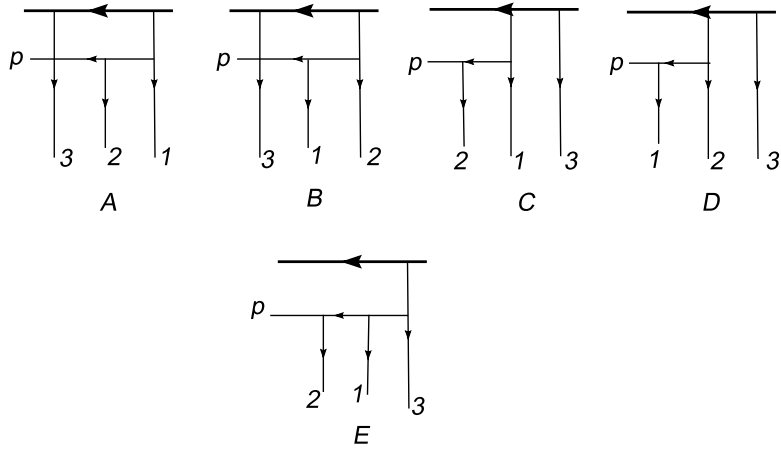}
\end{center}
\caption{Two and one interactions of the projectile with gluon emission}
\label{fig8}
\end{figure}

Taking into account that diagrams with 3-gluon interaction with the target give
zero, the total contribution from two interactions with the projectile reduces
to 4 diagrams shown in Fig \ref{fig8},$A-D$.
Diagrams B and D are obtained from A and C, respectively, by the
interchange $1 \leftrightarrow 2$. So we only need to study A and C.
The common momentum factor is
\[F_3= -16i\frac{(kl)^3 p_+}{2k_+^2}\frac{[e(p+q_1+q_2)]_\perp}{(p+q_1+q_2)^2+i0}\ .\]

The rest gives
\[(8.A)=F_3 \frac{(3d) f^{de1}f^{e42}}{((p+q_2)^2+i0)(3+i0)}\ ,\ \
(8.C)=F_3 \frac{(d3) f^{de1}f^{e42}}{((p+q_2)^2+i0)(-3+i0)}\ .\]

{\bf 4. Single interaction of the projectile}

Taking into account that diagrams with 4-gluon interaction with the
target give zero, the total contribution from the single interactions
with the projectile reduces to the diagram shown in Fig. \ref{fig8},$E$
plus others which are obtained by permutations of the three gluons
$1,2,3$.

The momentum factor is
\[F_4=16\frac{(kl)^3 p_+^2}{k_+^2}\frac{[e(p+q_1+q_2+q_3)]_\perp}
{(p+q_1+q_2+q_3)^2_\perp}\ .\]
Note that $(p+q_1+q_2+q_3)_-=0$ so that in this case
$(p+q_1+q_2+q_3)^2+i0=(p+q_1+q_2+q_3)^2_\perp$.
The rest gives the contribution from Fig. \ref{fig8},$E$ as
\beq
(8.E)=F_4\frac{t^bf^{bd3}f^{de1}f^{e42}}{((p+q_2)^2+i0)((p+q_2+q_1)^2+i0)}\ .
\label{c4}
\eeq

\subsection{Gluon emission off three centers in terms of
 the Lipatov and Bartels vertices}
To compare with the results of the effective action it is convenient to first
express the QCD amplitude in terms of the Lipatov and Bartels vertices.

We start from the expression for diagrams in Fig. \ref{fig7}
obtained in the previous subsection. Since it is symmetric under
permutations of gluons 1,2 and 3, we choose a different numeration
of gluons in different diagrams, namely, 213 from the left in the first
term, 132 in the second and 321 in the third one. Then instead of (\ref{eq10})
we get
$$
(7)=4\frac{(kl)^3}{k_+^2}\Big\{
\Big(\frac{(ep)_\perp}{p_\perp^2}-
\frac{[e(p+q_3)]_\perp}{(p+q_3)^2+i0}\Big)
\frac{(21d)f^{43d}}{(-(3+4)+i0)(2+i0)}
$$
$$
+\Big(\frac{(ep)_\perp}{p_\perp^2}-
\frac{[e(p+q_3)]_\perp}{(p+q_3)^2+i0}\Big)
\frac{(1d2)f^{43d}}{(-2+i0)(1+i0)}
$$
\begin{equation}
+\Big(\frac{(ep)_\perp}{p_\perp^2}-
\frac{[e(p+q_3)]_\perp}{(p+q_3)^2+i0}\Big)
\frac{(d21)f^{43d}}{(-1+i0)(3+4+i0)}\Big\} .
\label{eq13}
\end{equation}

We use the identity
\beq
\frac{1}{(p+q_3)^2 +i0}
=\frac{1}{(p+q_3)_{\perp}^2} -
\frac{2p_+ (p+q_3)_- }{(p+q_3)_{\perp}^2 ((p+q_3)^2 +i0)}\ .
\label{eq14}
\end{equation}
The first term in (\ref{eq14}) converts the brackets in (\ref{eq13})
into the Lipatov vertex and we get the first part of the amplitude as
\beq
A_1=(7)_1=4\frac{(kl)^3}{k_+^2}L(p,q_3)\Big\{
\frac{(21d)f^{a3d}}{(-(3+4)+i0)(2+i0)}
+
\frac{(1d2)f^{a3d}}{(-2+i0)(1+i0)}
\frac{(d21)f^{a3d}}{(-1+i0)(3+4+i0)}\Big\} +P_{123}\ .
\label{e131}
\end{equation}

The second term in (\ref{eq14}) cancels one of the denominators in
each of the three terms in (\ref{eq13}). For the second term it follows from
\begin{equation}
\frac{2p_+ (p+q_3)_- }{(-q_{2-}+i0)(q_{1-}+i0)}
=\frac{2p_+}{q_{1-}+i0} - \frac{2p_+}{-q_{2-}+i0}\ ,
\label{eq15}
\end{equation}
since $(p+q_3)_- = -(q_1 +q_2)_-$.
We obtain the second part of (7):
$$
(7)_2=4\frac{(kl)^3}{k_+^2}f^{43d}\cdot
\frac{[e(p+q_3)]_\perp}{(p+q_3)_{\perp}^2}
\frac{2p_+}{(p+q_3)^2+i0} \times
$$
\begin{equation}
\times
\Big\{ -\frac{(21d)}{2+i0}
+\frac{(1d2)}{1+i0}-\frac{(1d2)}{-2+i0}+\frac{(d21)}{-1+i0}\Big\}.
\label{eq16}
\end{equation}
Now we interchange $1\lra 2$ in the first and third terms in (\ref{eq16})
and change $d\to e$ after which it takes the form
\begin{equation}
(7)_2=8i\frac{(kl)^3}{k_+^2}\cdot p_+\frac{[e(p+q_3)]_\perp}{(p+q_3)_{\perp}^2}
\Big\{ \frac{(1d) f^{de2}f^{e43}}{((p+q_3)^2+i0)(1+i0)}
+ \frac{(d1)f^{de2}f^{e43}}{((p+q_3)^2+i0)(-1+i0)} \Big\}.
\label{eq18}
\end{equation}

Passing to contribution in Fig \ref{fig8},$A-D$ we choose the order of
gluons from left to right
$(132)$ for (8.A) and $(321)$ for (8.C). Then their sum can be rewritten as
\beq
(8.A+C)=-8i\frac{(kl)^3}{k_+^2}
\frac{[e(p+q_3+q_2)]_\perp}{(p+q_3+q_2)^2+i0}
\Big(
\frac{(1d) f^{de2}f^{e43} p_+}{(p+q_3)^2 +i0)(1+i0)}
+\frac{(d1) f^{de2}f^{e43} p_+}{(p+q_3)^2 +i0)(-1+i0)}
\Big) .
\label{eq17}
\eeq
Putting into (\ref{eq17}) the first term of the right-hand side of the identity
\begin{equation}
\frac{1}{(p+q_3+q_2)^2 +i0}
=\frac{1}{(p+q_3+q_2)_{\perp}^2} -
\frac{2p_+ (p+q_3+q_2)_- }{(p+q_3+q_2)_{\perp}^2 ((p+q_3+q_2)^2 +i0)}
\label{eq19}
\end{equation}
and summing with (\ref{eq18}) we get the second term of the amplitude
\begin{equation}
A_2=4\frac{(kl)^3}{k_+^2}\cdot i\frac{2p_+ B(p,q_3,q_2)}{(p+q_3)^2 +i0}
\left( \frac{(1d) f^{de2}f^{e43}}{1+i0}
+ \frac{(d1) f^{de2}f^{e43}}{-1+i0} \right) +P_{123} \ .
\label{eq101}
\end{equation}

Putting into (\ref{eq17}) the second term of the right-hand side
of (\ref{eq19}) we find
\begin{equation}
-16\frac{(kl)^3}{k_+^2}\frac{[e(p+q_3+q_2)]_\perp}{(p+q_3+q_2)_{\perp}^2}
\frac{t^b f^{bd1} f^{de2}f^{ea3}\cdot p_+^2}
 {((p+q_3)^2 +i0)((p+q_3+q_2)^2+i0)}\ .
\label{eq12}
\end{equation}
Summing this with the contribution (\ref{c4}) from Fig. \ref{fig8},$E$
we find the third term of the amplitude
\begin{equation}
A_3=-4\frac{(kl)^3}{k_+^2}B(p,q_3+q_2,q_1)
\frac{4p_+^2 \cdot t^b f^{bdb_1} f^{deb_2}f^{eab_3} }
 {((p+q_3)^2 +i0)((p+q_3+q_2)^2+i0)} +P_{123} \ .
\label{eq113}
\end{equation}

\begin{figure}
\hspace*{2.5 cm}
\begin{center}
\includegraphics[scale=0.75]{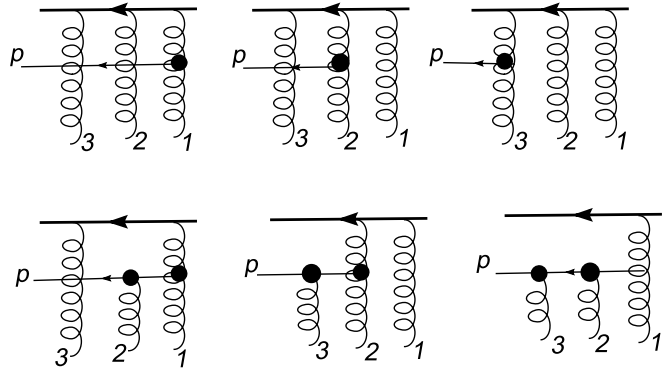}
\end{center}
\caption{Gluon emission on three centers in the lowest order of the QCD
presented in terms of Lipatov and Bartels vertices}
\label{fig9}
\end{figure}

As a result we presented the QCD amplitude as a sum of contributions
corresponding to the transverse picture with vertices of Lipatov and
Bartels with normal Feynman propagators both for gluons and quarks
and illustrated in Fig. \ref{fig9}.

\subsection{The effective action result}
To find the contribution for the amplitude following  application
of the effective action we can use our results in ~\cite{BSPV}
where the most complicated R$\to$RRRP vertex was constructed
under the assumption that neither of $q_{i-}$, $i=1,2,3$ vanishes.
It contributes to the part of the production amplitude $A_I$ with a single
interaction with the quark projectile.
The contribution $A_I$ was found to be
\footnote{Note that the results of ~\cite{BSPV} was rewritten here
in correspondence with the normalization $q_{\pm}=(q_{0}\pm q_{3})/\sqrt{2}$
accepted in the present paper.}
\begin{equation}
A_{I}=g^4  \gamma_{+}
t^{d}f^{d1 c}f^{c2d}f^{d3a}
\left(W_I + Q_I + R_I \right)
+ P_{123}\ ,
\label{e1a}
\end{equation}
where
\begin{equation}
W_I=\frac{2q_{+}^{2}B(p,q_{3}+q_{2},q_{1})}
{((q-q_{1})^{2}+i0)((q-q_{1}-q_{2})^{2}+i0)}\, ,
\label{e1w}
\end{equation}
\begin{equation}
Q_I=-\frac{q_{+}B(p,q_{3},q_{2})}
{q_{1-}((q-q_{1}-q_{2})^{2}+i0)}\, ,
\label{e1q}
\end{equation}
\begin{equation}
R_I=\frac{L(p,q_{3})}{2q_{1-}(q_{1-}+q_{2-})}\, .
\label{e1r}
\end{equation}
Here $q=p+q_1+q_2+q_3$; it is assumed that $q_{1-},(q_{1-}+q_{2-})\neq 0$.

In the framework of effective action to this contribution one has to add the
ones with the double and triple interactions of the projectile.
In both cases only the $\delta$-functional parts of the quark propagators
are to be retained. In this manner we find for the double interaction
with the projectile
\[
A_{II}=-i\pi g^{4}  \gamma_{+}
f^{d3a} \left\{ \Big((2d1)-(d21)+(12d)-(1d2)\Big)
\delta(1)\left[ \frac{q_{+}B(p,q_{3},q_{2})}{(q-q_{1}-q_{2})^{2}+i0}
-\frac{L(p,q_{3})}{2q_{2-}} \right]\right.\]
\beq
\left. +\delta(1+2)L(p,q_{3})
\left( \frac{(d21)}{2q_{1-}}
- \frac{(21d)}{2q_{2-}} \right)\right\}+P_{123}
\end{equation}
and for the triple interaction with the projectile
\beq
A_{III}=\frac{1}{2} g^{4}\gamma_{+}\pi^2\delta(1)\delta(2)
f^{d3a} \Big((d21)+(1d2)+(21d)\Big)
L(p,q_{3})+P_{123} \ .
\eeq

Summing all contributions we first find the term coming from $W_I$
in (\ref{e1w}) which exactly reproduces the QCD term $A_3$ (\ref{eq113}).
The part $Q_I$ summed with the term with $B(p,q_3,q_2)$ in $A_{II}$ gives
\[
g^{4}\gamma_{+}f^{db_3 a}
\frac{q_{+}B(p,q_{3},q_{2})}{(q-q_{1}-q_{2})^{2}+i0}\,
\Big[ \Big((2d1)-(d21) \Big)
\Big( -\frac{1}{q_{1-}} -i\pi\delta(q_{1-}) \Big)
\]
\[
+\Big((12d) -(1d2)\Big)
\Big( \frac{1}{q_{1-}} -i\pi\delta(q_{1-}) \Big) \Big ]
+ P_{123}\ .\]
We observe that this expression will coincide  with contribution
$A_2$ from the QCD provided we interpret the poles at $q_{1-}=0$
in the principal value sense.

The sum of all the rest terms can be presented as
\[
\frac{(2d1)_+-(d21)_+} {1(1+2)}
+\frac{(d21)_-+(1d2)_-}{q_{2-}}(-i\pi)\delta(1)
\]
\beq
+ \frac{(d21)_-}{q_{1-}}(-i\pi)\delta(1+2)
-\frac{1}{2}\Big((d21)_+ +(1d2)_+ +(21d)_+\Big)(-i\pi)^2 \delta(1)\delta(2)
+\Big(1\lra 2\Big)
\label{eq54}
\end{equation}
multiplied by factor $g^{4}\gamma_{+}f^{d3a}L(p,q_3)$.
Note that the first term comes from $R_I$ in the induced
$R\to RRRP$ vertex and has been obtained under the assumption
that $q_{1,2,3-}$, are different from zero, while the rest of the terms
involve contributions just from their zero values, coming from the
rescattering of the quark projectile.

In the sum with 1$\lra$2 in the first term there appears a contribution
\beq
(2d1)_+\Big(\frac{1}{1(1+2)}
+\frac{1}{2(1+2)}\Big).
\label{eq541}
\eeq
With the principal value poles the bracket is
\beq
\frac{1}{1(1+2)}+\frac{1}{2(1+2)}=\frac{1}{q_{1-}}\,\frac{1}{q_{2-}}
+\pi^2\delta(1)\delta(2).
\label{eq544}
\eeq
As we observe in this case the induced vertex again has a
$\delta$-functional contribution, which we must drop, according to our rule.
Then (\ref{eq541}) becomes
\beq
(2d1)_+\frac{1}{q_{1-}}\,\frac{1}{q_{2-}}
\label{e542}
\eeq
and one can combine all terms in (\ref{eq54}) into
\[
\frac{1}{2} (d21)
\Big(-\frac{1}{q_{1-}}\,\frac{1}{1+2}
+ (-i\pi)\delta(1) \frac{1}{1+2}
+ (-i\pi)\delta(1+2) \frac{1}{q_{1-}}
- (-i\pi)^2 \delta(1+2)\delta(1) \Big)
\]
\[
+ \frac{1}{2} (1d2)
\Big( \frac{1}{q_{1-}}\, \frac{1}{q_{2-}}
+ (-i\pi)\delta(1) \frac{1}{q_{2-}}
- (-i\pi)\delta(2) \frac{1}{q_{1-}}
- (-i\pi)^2 \delta(1)\delta(2) \Big)
\]
\beq
+ \frac{1}{2} (21d)
\Big( -\frac{1}{q_{2-}}\,\frac{1}{1+2}
- (-i\pi)\delta(2) \frac{1}{1+2}
- (-i\pi)\delta(1+2) \frac{1}{q_{2-}}
- (-i\pi)^2 \delta(1+2)\delta(2) \Big)
+ \Big(1\lra 2 \Big).
\label{e55}
\end{equation}
With all poles at $q_{1-}=0$, $q_{2-}=0$ and
$q_{1-}+q_{2-}=0$  taken in the principal value prescription
(\ref{e55}) can be written as
\beq
-\frac{1}{2}\left\{
 \frac{(d21)}
 { (-(1+2)+i0) (-1+i0) }
+\frac{(1d2)}
 {(1+i0) (-2+i0)}
+ \frac{(21d)}
 {(2+i0) (1+2+i0) } \right\}
+ \Big(1\lra 2\Big) .
\label{e56}
\end{equation}

Restoring the suppressed factor we see that this contribution exactly
coincides with the part $A_1$ of the QCD contribution, given by (\ref{e131}).
Thus, if we drop the $\delta$-functional term in (\ref{eq544}),
the effective action approach
give the correct QCD amplitude for gluon emission on three centers.

Note that this is not the only place where $\delta$-functional terms
appear in the induced vertex. In fact already the expression for the
R$\to$ RRRP vertex in $A_I$ was obtained after dropping such terms.
Calculations find $\delta$-functional terms in the induced R$\to$RRP vertex
\beq
\Delta \Gamma_{R\to RRP}=
-ig^2\pi^2\delta(1)\delta(2)
\left(p+q_1+q_2\right)_{\perp}^2\frac{(pe)_\perp}{2p_+}[b1a2+b2a1]+P_{12}
\eeq
and in the induced R$\to$RRRP vertex
\beq
\Delta \Gamma_{R\to RRRP}=
-g^3\pi^2 \delta(1)\delta(2) q_{\perp}^2 \frac{(pe)_\perp}{p_\perp^2}
\Big([bd21+b12d]f^{d3a}+ \frac{1}{2} f^{b1c}f^{c2d}f^{d3a}\Big).
\eeq
Collecting all
of them we find
that if they are retained the effective action result for the amplitude will
differ from the QCD by an extra term
\beq
g^4\frac{1}{8N_c}d^{2d1}f^{d3a}\gamma_+L(p,q_3)\pi^2\delta(1)\delta(2)
+P_{123}\ .
\eeq

\section{Discussion}
Two main results have been obtained in this paper.
First we have demonstrated that for collisions off two and three centers
in the Regge
kinematics high energy amplitudes both with and without gluon emission
can be presented in terms of the reggeon exchange with vertices of Lipatov
and Bartels for gluon emission with the standard Feynman propagators.
This greatly simplifies practical calculations of physical probabilities.
For the specific kinematical region $p_-<<q_{1,2-}$ this result was
already found earlier ~\cite{bralip}. In this paper we have demonstrated
that it it valid in the general Regge kinematics. Note that we believe that
this result has a more general validity and is true for any amplitudes.
This can be trivially demonstrated for the simple BFKL chain. Discussion
of more complicated examples can be found in ~\cite{braincl}.

Second we have formulated a rule which guarantees that the effective action
approach gives results coinciding with the QCD. This rule has the
complementarity property: crudely speaking, one has to retain only
the $\delta$-functional terms in the projectile propagators and to drop such
terms in the induced vertices, in which the poles at zero values
of the transferred longitudinal momenta are to be taken in the Cauchy principal
value sense. More precise meaning of the latter procedure was explained
in the Introduction.

Both results have been found in the lowest order of
the perturbative approach in the spirit of the BFKL approach,
in which higher orders are to be accompanied by evolution in rapidity.
They also have been found only for collisions on two and three centers.
We believe, however, that they preserve validity also for larger number
of collision centers.

As mentioned in Introduction a different approach was taken
in ~\cite{hent1,hent2}. M.Hentschinski proposed to project the induced
vertices for transitions of a reggeon to three or more reggeons
onto the maximally antisymmetric colour states and add $\pm i\epsilon$
to the denominators which vanish in these vertices. He found that after
this projection the dependence on the sign of $\epsilon$ vanishes
and the vertices satisfy the desired properties of Bose symmetry
and negative signature of the reggeon. It was checked that his recipe
reproduced the QCD results for the gluon trajectory with one and two
loops \cite{hent3,hent4,hent5}. However, his prescription refers only
to the induced vertices themselves and consequently to the amplitude
with a single interaction of colliding particles and not to the one
with several interactions as in our case. In the latter case
the amplitude may contain colour states different from those present
in the induced vertex after M.Hentschinski's projection.
We calculated the production amplitude off three centers with his
prescription for the vertices and only the $\delta$-functional parts
of the rescattering projectile propagators retained, as dictated
by the Regge kinematics. As a result we found an extra
(maximally symmetric) term compared to the QCD
\[\frac{1}{6}g^4\pi^2\gamma_+L(p,q_3)\delta(1)\delta(2)f^{d3a}
\Big([12d]_++[2d1]_++[d12]_+\Big) +P_{123} \ , \]
so that the prescription does not work.
Of course this result is valid only within our procedure to treat
multiple Regge exchanges (as in Fig. \ref{fig6},$A$), which we consider
well founded ~\cite{bralip}.
We admit that with a different procedure applied to Fig. \ref{fig6},$A$
the prescription of ~\cite{hent1,hent2} may work. However, to see that
one has first to propose an alternative procedure for Fig. \ref{fig6},$A$
compatible with the Regge kinematics.

\section{Acknowledgements}

The authors are most thankful to J.Bartels and G.P.Vacca for several
helpful and constructive discussions.
The authors acknowledge Saint-Petersburg State University
for a research grant 11.38.31.2011.
This work has been also supported by the RFFI grant 12-02-00356-a.
M.A.B. is indebted to the 2nd Institute
of theoretical Physics at Hamburg university
for hospitality.

\end{document}